\documentclass[universe,article,accept,pdftex,10pt,a4paper]{Definitions/mdpi} 

\firstpage{1} 
\pubvolume{4}
\issuenum{11}
\articlenumber{111}
\pubyear{2018}
\copyrightyear{2018}

\Title{Cosmic-Ray Extremely Distributed Observatory: status and perspectives}

\Author{D. G\'ora $^{1}$*\footnote{*Corresponding author: Dariusz.Gora@ifj.edu.pl} , K. Almeida Cheminant $^{1}$, D. Alvarez-Castillo$^{2}$, \L{}. Bratek $^{3}$, N. Dhital$^{1}$, A.R. Duffy $^{4}$, P. Homola $^{1}$, P. Jagoda $^{1,7}$,
J. Ja\l{}ocha $^{3}$, M.  Kasztelan $^{5}$,  K. Kopa\'nski $^{1}$, P.  Kovacs $^{6}$, V.  Nazari $^{1,2}$, M. Nied\'zwiecki $^{3}$, D. Ostrog\'orski $^{7}$, K. Rzecki$^{3}$,
K.  Smo\l{}ek $^{8}$, J. Stasielak $^{1}$, O. Sushchov $^{1}$, K. W. Wo\'zniak $^{1}$, J. Zamora-Saa $^{9}$ }

\AuthorNames{Dariusz Gora}

\address{%
$^{1}$ \quad Institute of Nuclear Physics Polish Academy of Sciences, Radzikowskiego 152,
Cracow, Poland\\
$^{2}$ \quad Joint Institute for Nuclear Research, Dubna, Russia \\
$^{3}$ \quad Faculty of Physics, Mathematics and Computer Science, Cracow University of Technology, Warszawska 24st 31-155 Cracow, Poland \\
$^{4}$\quad Centre for Astrophysics and Supercomputing, Swinburne University of Technology, Hawthorn, VIC 3122, Australia\\
$^{5}$ \quad National Centre for Nuclear Research, Andrzeja Soltana 7, 05-400 Otwock-Swierk, Poland\\
$^{6}$ \quad Institute for Particle and Nuclear Physics, Wigner Research Centre for Physics, Hungarian Academy of Sciences, H-1525 Budapest, Hungary\\
$^{7}$ \quad AGH University of Science and Technology, Cracow, Poland\\
$^{8}$ \quad Institute of Experimental and Applied Physics, Czech Technical University in Prague\\
$^{9}$ \quad Universidad Andres Bello, Departamento de Ciencias Fisicas, Facultad de Ciencias Exactas, Avenida Republica 498, Santiago, Chile\\
}

\corres{Correspondence: Dariusz.Gora@ifj.edu.pl}

\abstract{The Cosmic-Ray Extremely Distributed Observatory (CREDO) is a project dedicated to global studies of extremely extended cosmic-ray phenomena, the cosmic-ray ensembles (CRE), beyond the capabilities of existing detectors and observatories. Up to date cosmic-ray research has been focused on detecting single air showers, while the search for ensembles of cosmic-rays, which may overspread a significant fraction of the Earth, is a scientific terra incognita. Instead of developing and commissioning a completely new global detector infrastructure, CREDO proposes approaching the global cosmic-ray analysis objectives with all types of available detectors, from professional to pocket size, merged into a worldwide network. With such a network it is possible to search for evidences of correlated cosmic-ray ensembles. One of the observables that can be investigated in CREDO is a number of spatially isolated events collected in a small time window which could shed light on fundamental physics issues. The CREDO mission and strategy requires active engagement of a large number of participants, also non-experts, who will contribute to the project by using common electronic devices (e.g. smartphones). In this note the status  and perspectives of the project is presented.}

\begin{document}


\section{Introduction}
\label{intro}

Considering the problem of ultra-high energy cosmic-rays (UHECR) origin, one should keep in mind that no ultra-high energy (UHE) photons (above $10^{18}$ eV) have been observed even by the largest observatories so far~\cite{auger_photons}. It is possible that there is no mechanism of acceleration of photons with higher energies. Alternatively, as predicted by Lorentz invariance violation (LIV) models~\cite{LIV}, photons may decay after a very short time (of the order of 1 second). In this case, the direct observation of such a photon at Earth seems to be impossible, because they decay nearly immediately and thus have no chance to reach Earth. On the other hand, it is believed that cascades initiated by such photons, are totally dissipated on their way to Earth, although no precise calculations of the horizon (the distance within which a cascade can reach Earth) with different theoretical assumptions have been made so far. There are  however  the  so-called  bottom-up (astrophysical), or  top-down (exotic) scenarios which predict the presence of UHE photons in the cosmic-ray flux~\cite{photons}.

 The possible explanations of absence of UHE photons could be  related with the fact that existing techniques are not capable of distinguishing hadronic shower and CRE.   One can also speculate that UHE photons do not reach the Earth because of some yet unknown or not well understood processes occurring before entering the Earth's atmosphere. As a result of those interactions electromagnetic cascades may be developing. Since such processes may occur at very large distances from the Earth, the transverse size of the mentioned cascades can be comparable with the Earth's diameter.

The main objective of the CREDO is detection and performing global analysis of CRE,
 using all accessible infrastructure~\cite{alex}. Unlike the state of-the-art approach, oriented at the detection of single cosmic-ray events,
  the search for CRE is yet unprobed. The key idea of CREDO is collecting existing cosmic-ray detectors (large professional arrays, educational instruments, individual detectors, such as smartphones, etc.) in a worldwide network, thus constructing a scientific tool for global analysis. The second goal of CREDO is involving large number of participants (citizen
science!), assuring geographical spread and managing manpower, necessary to deal
with the vast amount of data.
\begin{figure}[h]
\centering
\includegraphics[width=10cm,clip]{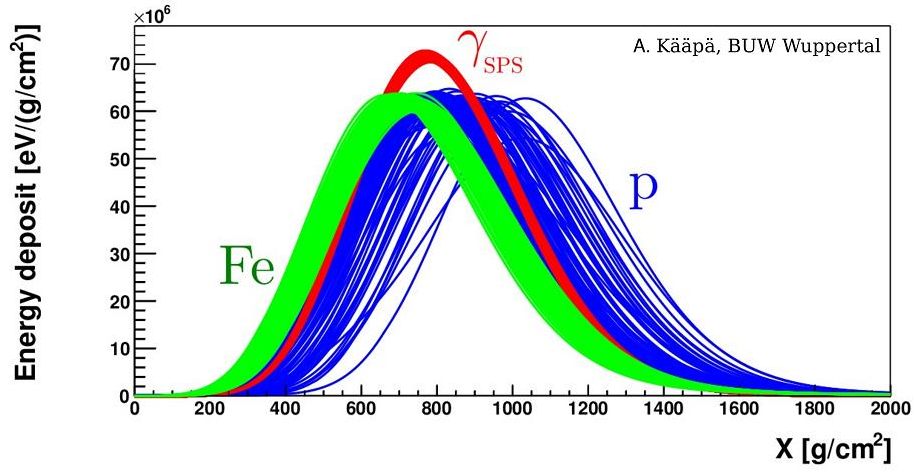}
\caption{Longitudidal development of air showers induced by protons, iron nuclei 
and super-preshowers of primary energies $E=10^{19.6}$ eV, arriving at the Pierre Auger 
Observatory~\cite{pao} from geographical South, at zenith angles of 60 deg~\cite{xmax}. }
\label{fig-xmax}       
\end{figure}

\section{The Sun and Earth  preshowers}
\label{sec-1}

A good candidate for a  CRE is a shower   induced by ultra high energy photon  interacting close by the Sun or the Earth.  Such cascades, which we call super-preshowers (SPSs), are produced as a consequence of interaction of UHE photons with solar magnetic field. Chain of SPS development involves pair production of leptons (predominantly $e^{+} e^{-}$) if the magnetic field component transverse to the direction of propagation of primary UHE photon is large enough. For primary energies greater than $10^{20}$ eV, typical solar magnetic field at a distance of a few solar radii from the Sun is sufficient to initiate the process. Furthermore, as leptons thus produced propagate through the magnetic field, they emit synchrotron photons. Among these photons, the ones which have very high energies can undergo pair production again. Consequently, by the time the cascade arrives at the Earth, it comprises several thousand photons and a few leptons with a peculiar spatial distribution to which CREDO will be very sensitive ~\cite{homolaSPS}. In reality, electromagnetic cascading process of UHE photons occurs even in the Earth’s vicinity~\cite{preshower,preshower2}. An event-by-event identification of preshowers in cosmic-ray observatories, however, is limited by the fact that the constituent particles in a preshower have very narrow transverse  spread at the top of atmosphere, typically of the order of a few centimetres. An important consequence of preshower phenomenon is a decrease in $X_{max}$, the atmospheric depth at which the number of particles in an extensive air showers (EAS) reaches its maximum, for UHE photon initiated EASs towards that for proton initiated EASs. In Figure~\ref{fig-xmax} the position of the maximum, $X_{max}$, of the energy deposit as a function of $X$ for a photon SPS is within the range of $X_{max}$, values observed for protons and nuclei, only the width of the function is slightly smaller. Since $X_{max}$ is one of main parameters used in UHECR research,  it is prudent to perform the UHE photon search using an alternative approach for a better inference on the mass composition of UHECRs and other physics which derived thereof. CREDO will implement search strategies aimed at SPS detection to serve this purpose.
\begin{figure}[t]
\centering
\includegraphics[width=7cm,height=5.5cm,clip]{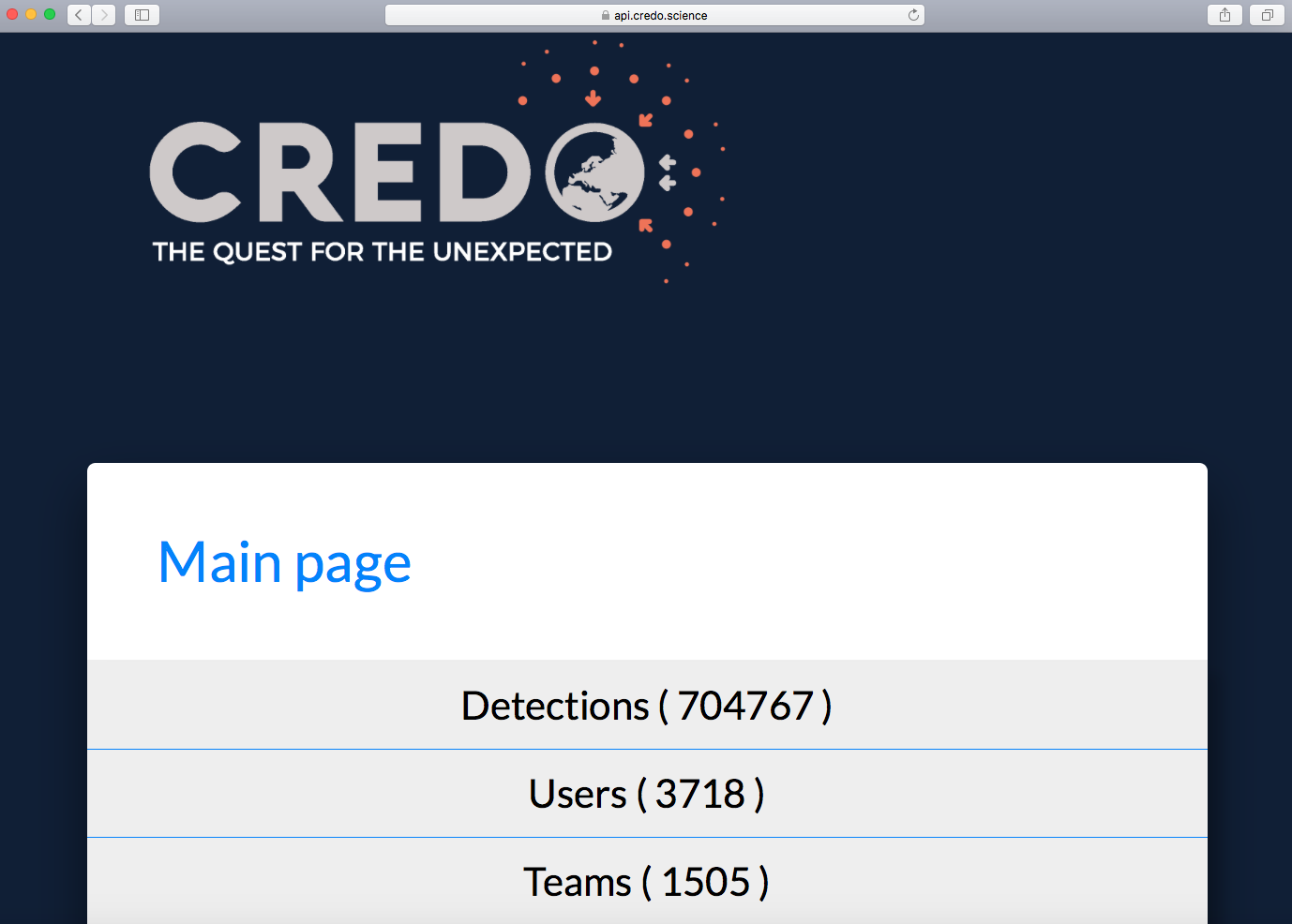}
\includegraphics[width=8cm,height=5.5cm,clip]{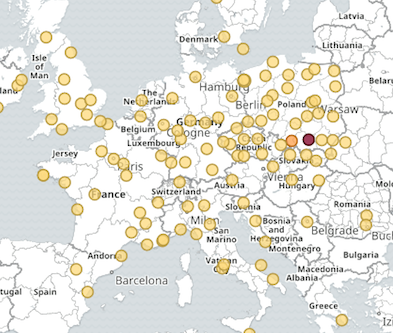}
\caption{The API server of CREDO with the number of user and detections (left)  and the map of CREDO user locations as of middle October 2018 (right). 
}
\label{fig-users}       
\end{figure}
 

\section{Smartphones  as cosmic-ray detector}
\label{sec-2}
   One of the key ideas to make CREDO as large as possible is to involve in the experiment also non-scientists and their pocket devices: smartphones. The idea of using a smartphone as a portable   cosmic-ray   detector   has   been   practically   explored   by   two   collaborations:   Distributed Electronic Cosmic-
ray  Observatory (DECO)~\cite{deco}   and  Cosmic-RAYs Found In Smartphones (CRAYFIS) ~\cite{crayfis}~\footnote{The DECO collaboration has been demonstrated discrimination between GeV cosmic-ray muon tracks and MeV electron tracks caused by radioactive decay using morphological cuts based on
track images~\cite{decoana}. However, large number  of smartphones ($\sim 10^{6}$~\cite{unger}) are needed to reach the sensitivity  comparable to the largest  cosmic-rays observatories like the  Pierre Auger Observatory or Telescope Array. In order to obtain valuable results it is thus necessary to combine the data from smartphones with the data from other detectors.}. DECO offers a free cosmic-ray-detection application already while  CRAYFIS is still in the beta phase. Recently  CREDO has also   developed  a free application to detect the cosmic-rays~\cite{credoapp}. The  CREDO Detector application for Android smartphones can be downloaded from the official Google Play store, from the IFJ PAN account. After installation, one has to register the device  in the CREDO system. The CREDO Detector uses the image detector in the smartphone's camera. If the camera's lens is obscured, the images recorded should consist of only black pixels. When this condition is met, the CREDO Detector starts working.
If a particle of secondary cosmic radiation, or possibly a particle of a local radiation, passes through the camera in the smartphone, some of the pixels of the sensor should register a signal.  A few to several dozen lighter pixels, grouped in a more or less fanciful shape, sometimes circular, sometimes stretched into a line,  appears on the homogeneous black background. In a 24-hour period from a few to several hundred detections are expected.  Images are stored on a central database for future analysis, and can be browsed using the CREDO web interface~\cite{apicredo} as shown in Figure~\ref{fig-users}.  As we can see 
 there the number of registered users in the middle of October 2018 was 3718 and about 704767 images were stored in database. In general, there are three distinct types of charged particle events seen by CREDO Detector application: tracks, worms, and
spots.  These are named according to the convention in~\cite{groom},  which categorizes events by the morphology of their resulting images. Tracks are straight, likely due to high energy (GeV
scale) cosmic-ray muons that exhibit little Coulomb scattering.  Worms show significant deflection along the trajectory, curving and/or kinking. They are most likely due to low-energy (MeV scale) electrons what exhibit multiple Coulomb scattering. Such electrons are a product of radioactive decay in the materials of the phone itself or in the surroundings.
The source of these low-energy electrons is typically the Compton scattering of gamma rays from potassium-40, and from uranium or thorium isotopes decay chains. Much smaller, diffusion-limited hits are also observed (spots), and are generally attributed to nuclear recoil events, very low-energy Compton scatters (short worms), or low energy X-rays.
\begin{figure}[t]
\centering
\includegraphics[width=10cm,clip]{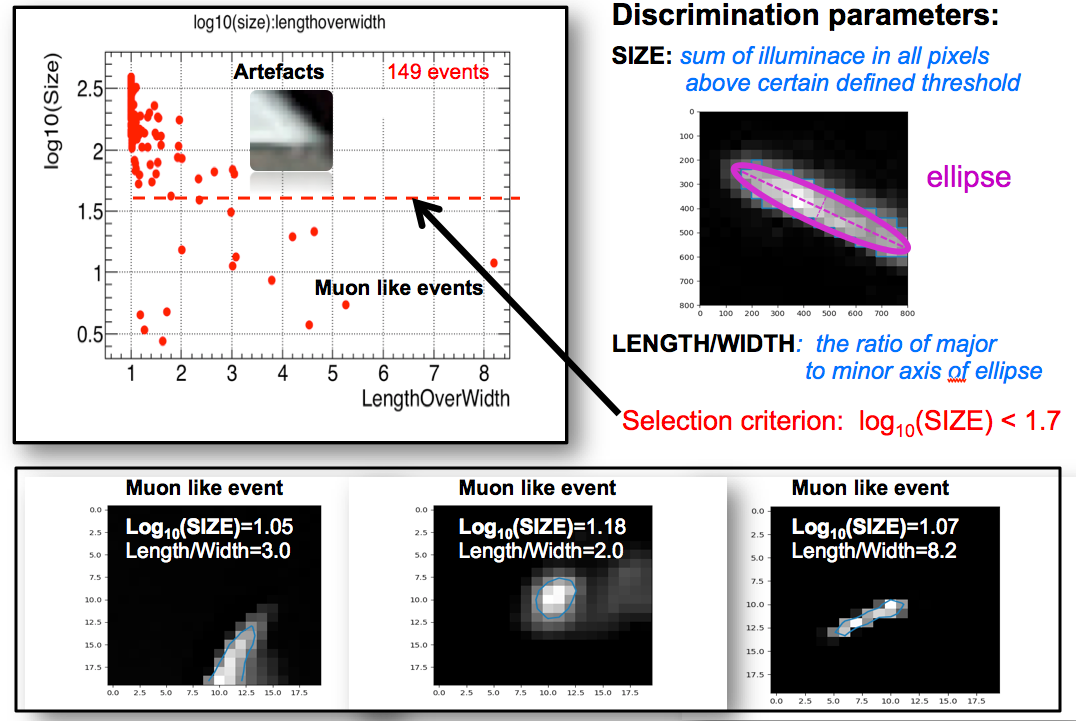}
\caption{Example of clasification of CREDO smartphone's images (see  text for more details). }
\label{fig-filter}       
\end{figure}
To classify events and to extract tracks (muon-like events) we can use the method from~\cite{decoana}. Thus,  in a  first step,  we convert the RGB pixel values to a single gray-scale amplitude (luminance) using the formula from~\cite{rgb}. Then  we calculate a contour using the “marching
squares” algorithm implemented in python module Scikit~\cite{scikit}, to delimit the pattern of pixels that detected ionization above a particular threshold. Finally, several observables are calculated for each event like  the total luminance above threshold (defined here  as the SIZE parameter)
or  the Length and Width parameter  of ellipse obtained from Principal Component Analysis (PCA)~\cite{pca}. 
An  example  of  simple analysis for some  preliminary data set of  images from smartphones and using a simple selection criterion is shown in Figure~\ref{fig-filter}. As we can see from the plot track events  have moderate SIZE parameter value: $\log_{10}(SIZE) < 1.7$.

\begin{figure}[t]
\centering
\includegraphics[width=7cm,height=7.5cm, clip]{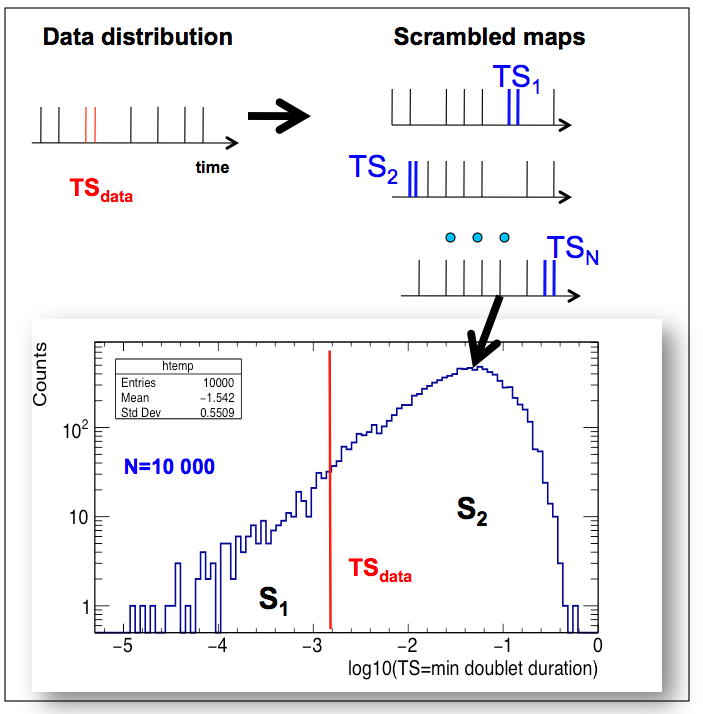}
\includegraphics[width=7cm,height=7.5 cm,clip]{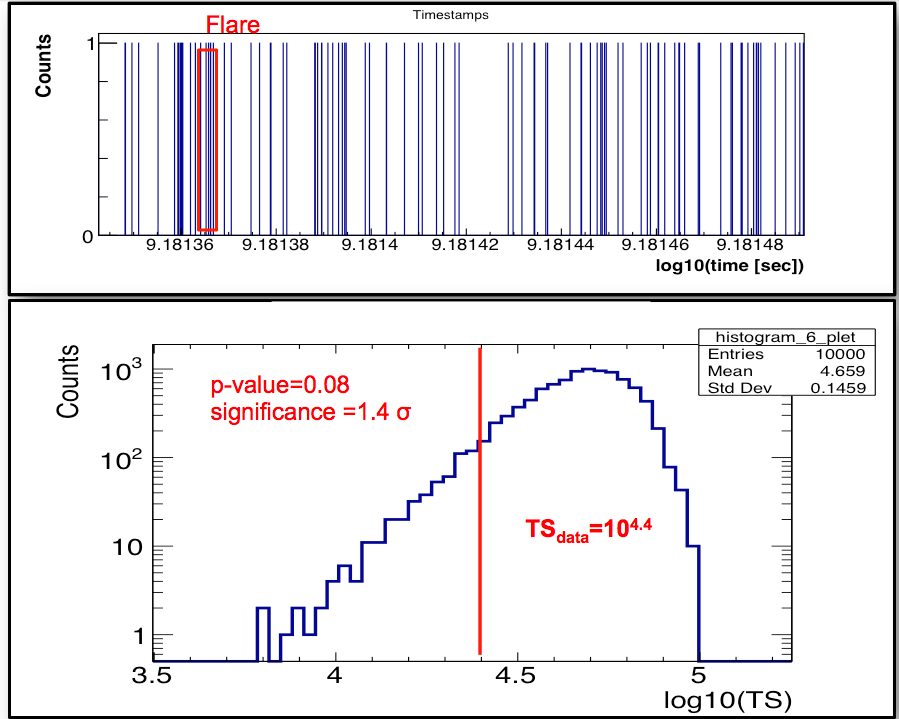}
\caption{Left panel: Graphical explanation of time clustering algorithm.  For doublets the test statistic (TS)  is defined as the minimum time difference between  two consecutive events. The test statistic value for the data  TS$_{data}$ was shown as the vertical red line. The p-value can be done by obtaining the ratio of the area on the left  of the red line ($S_{1}$) over the total area of the distribution (S$_1$+S$_2$). This p-value can be converted into a sigma  by using the error  function of the  Gauss distribution,  see text for more details.  Right panel: Distribution of 100  detection timestamps from one smartphone corresponding to about 6 days (First event: 10.02.2018 at 4 h 26 m 6 s;  Last event: 16.02.2018 at 7 h 38 m 56 s) of taking data with the CREDO Detector application. Some excess seems to appear at the beginning of the data acquisition i.e cluster of 6 events (10.02.2018 22 h 49 m 35 s - 11.02.2018 00 h 36 m 16 s)  with significance of about 1.4 $\sigma$ but this result is compatible with the background expectation.}
\label{fig-ts}       
\end{figure}

The aim of CREDO is also to look for time coincidences between the cosmic-ray signal received by very distant stations, on a globe scale. 
One of the methods for a time  search is to look for events coming from a restricted angular region, which could be identified with a known astrophysical object. For such a source in the sky an excess of events from a particular direction over the background is expected.
 These events might present additional features that distinguish them from background, for example a different energy spectrum or formation of  a short time cluster in time. In such search we would like to use not only the data from smartphones but also publicly available data from other type of detectors (cosmic-ray detector arrays, CCD camera image from astrophysical data, etc.). The   search for the time clusters  is going to be performed with methods presented in ~\cite{gora2011}, but with the time probability functions only. Essentially, the time clustering algorithm selects the most significant cluster of events in time and returns the mean time and width of the corresponding flare.  In a first step, the method selects the most promising flare candidates over different time windows, which are given by the combination of the times of signal-like events from the analysed data set. 
 Each pair of events with times $t_i$  and $t_j$  defines a candidate flare time window ($\Delta t_{i,j}$). For each $\Delta t_{ij}$, a significance parameter (the test statistic) $TS_{i}$ is then calculated as defined in~\cite{gora2011}.  Larger values of $TS_{i}$ correspond to data less compatible with the null hypothesis (i.e. zero expected signal events in the data sample tested). Finally, the algorithm returns the best $TS_{i}$, $TS_{data}$, corresponding to the most significant time window over the entire data period analysed.  Significance of a cluster of chosen clusters can be determined by comparing the highest $TS$ with the distribution obtained for data sets scrambled in time, see Figure~\ref{fig-ts} for explanation  of the algorithm and an example of data analysis. 

\section{Outreach and citizien science}
The CREDO project needs the help from citizens, thus the communication with the general public is
also crucial. We can achieve this using several channels which include: Dark Universe project~\cite{darkuniverse} -  attracting most enthusiastic persons willing to help CREDO; Android application changing smartphones into detectors; CREDO web page~\cite{credoweb} with general information on the project, public lectures, interviews, instructions both written and in video format (youtube), possibly in several languages.

The  idea is to explore smartphone capability of being a particle detector to induce and drive
scientific enthusiasm of the public, then cultivate this enthusiasm by providing a range of
opportunities of undertaking a long term scientific journey towards personal development. The first step is API service hosted by ACC Cyfronet AGH-UST~\cite{cyfronet}. Everyone is encouraged to produce, using our open API server, his  own (but accessible and open to community) web solutions and applications to visualize and analyse the data, including customised rankings and statistics. We plan also to maintain and develop central web application with all  basic funcionalities, enabling e.g. easy following and customizing views 
of the monitors of the CREDO on-line experiments. All the codes maintained and developed by the CREDO is open with the MIT licence~\cite{mit},  increase the scale, dynamize the engagement. The general philosophy is: to break the producer-receiver
relation (like a car factory and its customers) and change it into “we build the project together“. The prototypes of the key software ingredients already exist: the mobile application CREDO Detector and API server are available as MIT licensed codes on the GitHub~\cite{credo-github}. We have also developed a prototype of the citizen science workflow to work within CREDO i.e.  Dark Universe Welcome ~\cite{darkuniverse}. This  workflow is  located on the
zooniverse platform where participation of external developers is limited, the ideas behind the workflow are transferable to other existing citizen science platforms with open source – thus our current prototypes can be considered showcases encouraging others to build clones with open source
solutions to increase flexibility and benefit from contributions of external developers.

\section{Summary}
Pursuing the research strategy proposed in CREDO will have large impact on astroparticle physics
and possibly also on fundamental physics. If CRE are found, they could point back to the interactions at energies close to the   Grand Unified Theories (GUT)
scale. This would give  an unprecedented chance to test experimentally dark matter models and scenarios,
probe the interaction models, the properties of  spacetime itself – all in the close to-GUT energy regime. If CRE are not observed it would valuably constrain the current and future  theories.
Apart from addressing fundamental physics questions the CREDO has a number of
additional applications: alerting the astroparticle community on CRE candidates to enable a multi-channel
data scan, integrating the scientific community
(variety of science goals, detection techniques, wide cosmic-ray energy ranges, etc.), helping non-scientists to explore  Nature on a fundamental but still understandable level – and many more.  The high social and educational potential of the project    gives confidence in its contributing to a progress in physics.

\funding{This work was partly funded by the International Visegrad Fund under the grant no. 21720040.}

\acknowledgments{This research has been supported in part by PLGrid Infrastructure.  We warmly thank the staff at
ACC Cyfronet AGH-UST, for their always helpful supercomputing support. CREDO application is developed in Cracow University of Technology.
}

\conflictsofinterest{The authors declare no conflict of interest.} 

\abbreviations{The following abbreviations are used in this manuscript:\\

\noindent 
\begin{tabular}{@{}ll}
CREDO & The Cosmic-Ray Extremely Distributed Observatory\\
CRE & cosmic-ray ensembles\\
UHECR& ultra-high energy cosmic-rays \\
SPS &super-preshower\\
\end{tabular}}



\reftitle{References}





\end{document}